\documentclass[%
reprint,
superscriptaddress,
amsmath,amssymb,amsthm,mathrsfs,
aps,
prl,
]{revtex4-2}
\usepackage{float}
\usepackage{graphicx}
\usepackage{dcolumn}
\usepackage{url}
\usepackage{bm}
\usepackage[usenames]{color}
\usepackage[ 
margin=0.75in,
]{geometry}
\usepackage[normalem]{ulem}

\def\nyo{NaYbO$_2$}

\def\onetwenty{120$^{\circ}$}

\begin{document}
	
\title{
	Quantitative investigation of the short-range magnetic correlations in candidate quantum spin liquid NaYbO$_2$
}
	
	\author{Kristina Michelle Nuttall}
	\affiliation{ %
		Department of Physics and Astronomy, Brigham Young University, Provo, Utah 84602, USA.
	} %

	\author{Christiana Z. Suggs}
	\affiliation{ %
		Department of Physics and Astronomy, Brigham Young University, Provo, Utah 84602, USA.
	} %

\author{Henry E. Fischer}
\affiliation{Institut Laue-Langevin, BP 156, 38042 Grenoble Cedex 9, France}

	\author{Mitchell M. Bordelon}
    \affiliation{Materials Department, University of California Santa Barbara, Santa Barbara, California 93106, USA}

	\author{Stephen D. Wilson}
    \affiliation{Materials Department, University of California Santa Barbara, Santa Barbara, California 93106, USA}

 \author{Benjamin A. Frandsen}
	\affiliation{ %
		Department of Physics and Astronomy, Brigham Young University, Provo, Utah 84602, USA.
	} %

\begin{abstract}
We present a neutron diffraction study of \nyo, a candidate quantum spin liquid compound hosting a geometrically frustrated triangular lattice of magnetic Yb$^{3+}$ ions. We observe diffuse magnetic scattering that persists to at least 20~K, demonstrating the presence of short-range magnetic correlations in this system up to a relatively high energy scale. Using reverse Monte Carlo and magnetic pair distribution function analysis, we confirm the predominant antiferromagnetic nature of these correlations and show that the diffuse scattering data can be well described by noninteracting layers of Heisenberg or XY spins on the triangular lattice. We rule out Ising spins and short-range-ordered stripe or \onetwenty\ phases as candidate ground states of \nyo. These results are consistent with a possible QSL ground state in \nyo\ and showcase the benefit of combined reciprocal- and real-space analysis of materials with short-range magnetic correlations.
\end{abstract}
	
\maketitle

\section{Introduction}	
Antiferromagnetically coupled $S=1/2$ spins on a triangular lattice constitute a prototypical architecture for exotic quantum magnetism induced by geometric frustration, including highly entangled ground states such as quantum spin liquids (QSLs)~\cite{ander;mrb73,balen;n10,li;jpcm20}. Although they are growing in number, real triangular lattice antiferromagnets that can be considered genuine QSL candidates are still quite rare~\cite{li;jpcm20,savar;rpp17,yi;rmp17,broho;s20}. 
In that context, the recent discovery of a large family of rare-earth chalcogenides with the formula $ARX_2$ ($A$ = alkali or monovalent cation, $R$ = rare earth, $X$ = O, S, or Se)~\cite{hashi;jssc03,liu;cpl18} is significant. These compounds crystallize into a simple delafossite-type structure with space group $R\overline{3}m$ in which the magnetic rare earth ions form planes of perfect triangular lattices. They are free of chemical disorder, thereby avoiding complications such as the Mg/Ga site mixing in the triangular lattice antiferromagnet YbMgGaO$_4$~\cite{paddi;np17,li;aqt19}. A flurry of recent research activity has already revealed strong evidence for quantum disordered ground states in numerous compounds in this family~\cite{baeni;prb18,ding;prb19,ranji;prb19,ranji;prb19b,borde;np19,xing;prb19,borde;prb20,guo;prm20,schei;prb20,xing;acsml20,dai;prx21,ortiz;np23}, raising the exciting possibility of studying QSL physics in this diverse class of materials.

\begin{figure}
	\includegraphics[width=80mm]{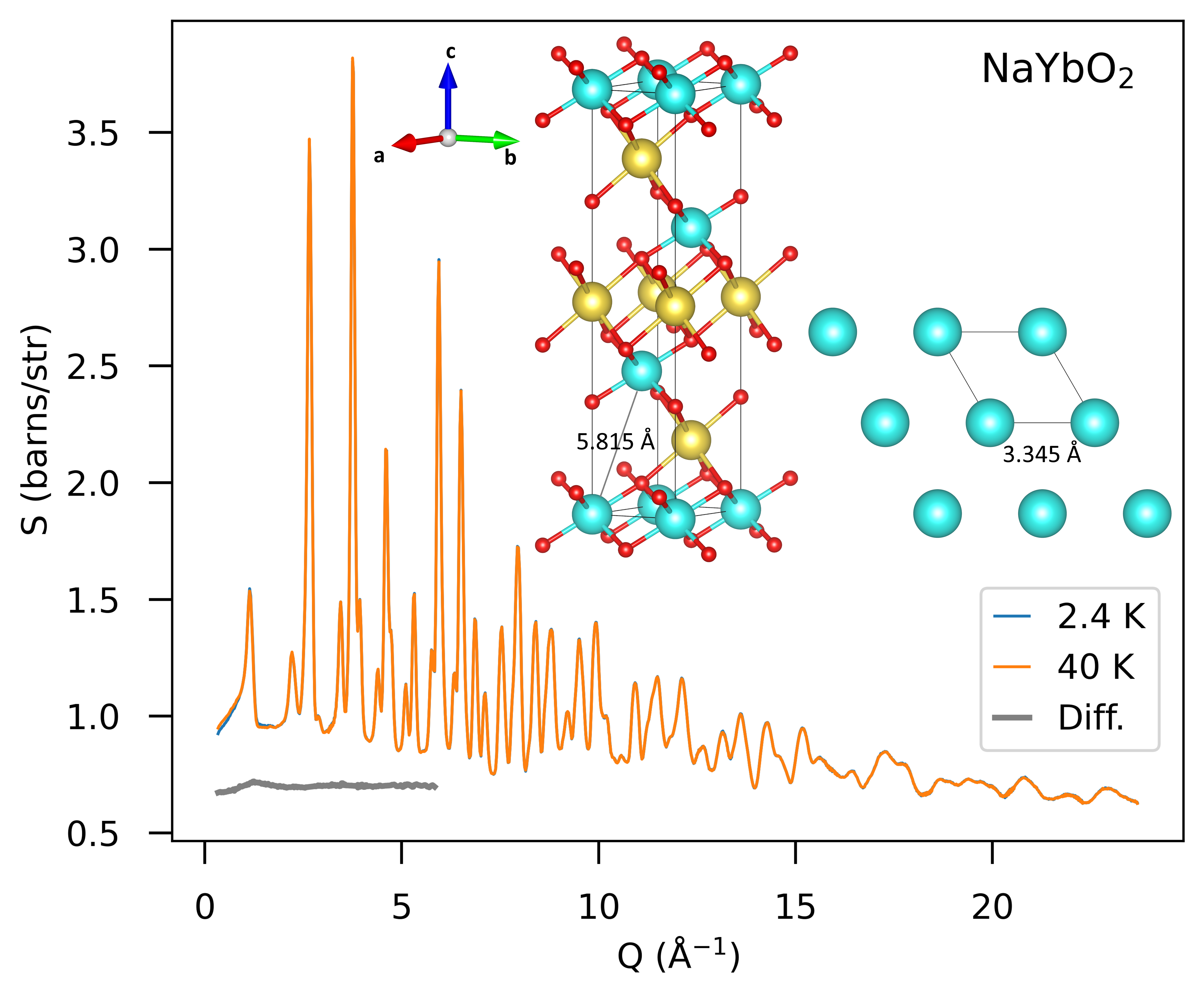}
	\caption{\label{fig:diff-struc} Neutron total scattering structure function of \nyo\ at 2.4~K and 40~K. The difference (i.e. subtracting the 40-K intensity from the 2.4-K intensity) up to 5.85~\AA$^{-1}$ is offset below in gray. Inset: Crystal structure of \nyo, including a top-down view of the triangular lattice formed within each layer of Yb atoms. The distances labeled 3.345~\AA\ and 5.815~\AA\ are the shortest in-plane and out-of-plane Yb-Yb distances, respectively.}
\end{figure}

Among the most promising of these new materials is \nyo. The crystal structure, shown in the inset of Fig.~\ref{fig:diff-struc}, features a perfect triangular lattice of $J_{\mathrm{eff}}=1/2$ moments localized on the Yb$^{3+}$ ions~\cite{liu;cpl18,borde;np19}.  No magnetic long-range order or spin freezing has been detected at any temperature down to $\sim$50~mK, while signatures of fluctuating short-range correlations are observed by magnetometry, heat capacity, nuclear magnetic resonance, and muon spin spectroscopy up to approximately 20~K~\cite{borde;np19,ding;prb19,ranji;prb19}. Inelastic neutron scattering experiments have revealed a gapless continuum of spin excitations at low temperature~\cite{borde;prb20}. These experimental observations are all consistent with a QSL ground state in \nyo. 

Extensive theoretical work has been done on the triangular lattice. For isotropic Heisenberg $J_1 - J_2$ models, a range of possible ground states has been predicted depending on the strength of the exchange interactions, including  \onetwenty\ order, stripe antiferromagnetism, noncoplanar tetrahedral order, and QSLs~\cite{zhu;prb15,hu;prb15,saada;prb16,iqbal;prb16,gong;prb17,hu;prl19,}. For \nyo\ and related materials like YbMgGaO$_4$, models with anisotropic exchange are also relevant~\cite{paddi;np17,borde;np19}. For these models, expectations for a QSL ground state are mixed: one study using the density-matrix renormalization group predicts a small QSL region in the phase diagram~\cite{zhu;prl18}, while another using the projected entangled pair state method does not~\cite{zhang;prb22}. 

Detailed knowledge of the short-range magnetic correlations in \nyo\ is essential for gaining a better understanding of the ground state of this compound. Analysis of diffuse neutron scattering arising from short-range magnetism is one of the few experimental tools available to gain this type of information. The diffuse scattering may be analyzed in either reciprocal space or real space via the Fourier transform. The latter, referred to as the magnetic pair distribution function (mPDF), yields the pairwise local magnetic correlations in real space~\cite{frand;aca14,frand;aca15}. Detailed modeling of the data can provide quantitative information about the short-range correlations, e.g. via reverse Monte Carlo (RMC) refinements in which spins in a large supercell are randomly adjusted until agreement with the data is achieved~\cite{paddi;prl12} (big-box modeling), or through more constrained refinements assuming a magnetic unit cell with a finite correlation length (small-box modeling). Magnetic RMC and mPDF analysis have been applied successfully to correlated paramagnetic states in numerous types of materials, including geometrically frustrated magnets~\cite{paddi;np17, paddi;prl13, paddi;prb14, paddi;s15, frand;prl16, paddi;nc16, frand;prm17,lefra;prb19, frand;prb20,dun;prb21,baral;matter22}.

Here, we present a detailed analysis of the diffuse magnetic scattering and corresponding real-space mPDF data of \nyo\ at temperatures down to 2.4~K, low enough to observe the short-range magnetic correlations inferred from magnetometry, heat capacity, and muon spin spectroscopy data~\cite{borde;np19,ding;prb19,ranji;prb19}. The results reveal dominant NN antiferromagnetic correlations that persist up to at least 20~K, with further neighbor correlations up to approximately 1~nm present at 2.4~K. The data can be accurately reproduced through RMC refinements of a single layer of Yb$^{3+}$ ions occupying a triangular lattice decorated with Heisenberg or XY spins oriented within the plane of the triangular lattice. Ising spins provide a worse fit to the data, as do short-range-ordered \onetwenty\ and stripe antiferromagnetic spin configurations. These findings offer a substantially more detailed view of the magnetic ground state in \nyo\ than was previously available, further establishing the possibility of a genuine QSL ground state in \nyo\ and providing important constraints for future theoretical investigations of this material.

\section{Methods}
A powder sample of \nyo\ of mass 5.57~g was synthesized via solid state reaction as described in Ref.~\onlinecite{borde;np19}. Laboratory x-ray diffraction and magnetometry measurements are consistent with previous results~\cite{borde;np19, borde;prb20}, confirming the good quality of the sample.

Neutron total scattering experiments~\cite{fisch;data20} were performed using the hot neutron diffractometer D4~\cite{fisch;apa02} at the Institut Laue-Langevin with a constant wavelength beam ($\lambda = 0.5$~\AA). The sample was loaded into a cylindrical vanadium can with outer diameter 7~mm and placed in a cryofurnace with a base temperature of 2.4~K. The sample height in the can was 50~mm. Total scattering patterns were collected at several temperatures between 2.4~K and 40~K, reduced according to standard protocols at D4 (including attenuation and multiple-scattering corrections), and normalized by a vanadium standard to obtain the total scattering structure function $S(Q)$ in absolute units. The diffraction pattern collected at 40~K was subtracted from the patterns collected at lower temperatures to remove the nuclear Bragg peaks and isolate the diffuse magnetic scattering. Small temperature-dependent shifts in the Bragg peak positions due to thermal expansion were corrected by manual interpolation through the characteristic ``S''-shaped features resulting from the imperfect subtraction. The momentum transfer range included in our analysis was 0.34 to 5.85~\AA$^{-1}$. The mPDF was generated via Fourier transformation with $Q_{\mathrm{max}}=5.85$~\AA\ and a Fermi-Dirac modification function~\cite{frand;jap22} of width 1~\AA$^{-1}$ using \texttt{diffpy.mpdf}~\cite{frand;jac22}. Because the detectors at D4 effectively integrate over all energy transfers, the diffuse magnetic scattering probes the instantaneous spin-spin correlations in \nyo.

SPINVERT and SPINCORREL~\cite{paddi;jpcm13} were used to perform RMC modeling of the data in reciprocal space and extract the spin correlation function from the RMC-generated spin configurations. \texttt{diffpy.mpdf} was used to generate the mPDF data from the diffraction data and conduct small-box modeling of the mPDF data up to 10~\AA\ in real space. We used the analytical approximation to the magnetic form factor of Yb$^{3+}$ provided in~\cite{itable;volc95} with a Land\'e splitting factor of $g=8/7$.

\section{Results and Discussion}

\begin{figure*}
	\includegraphics[width=160mm]{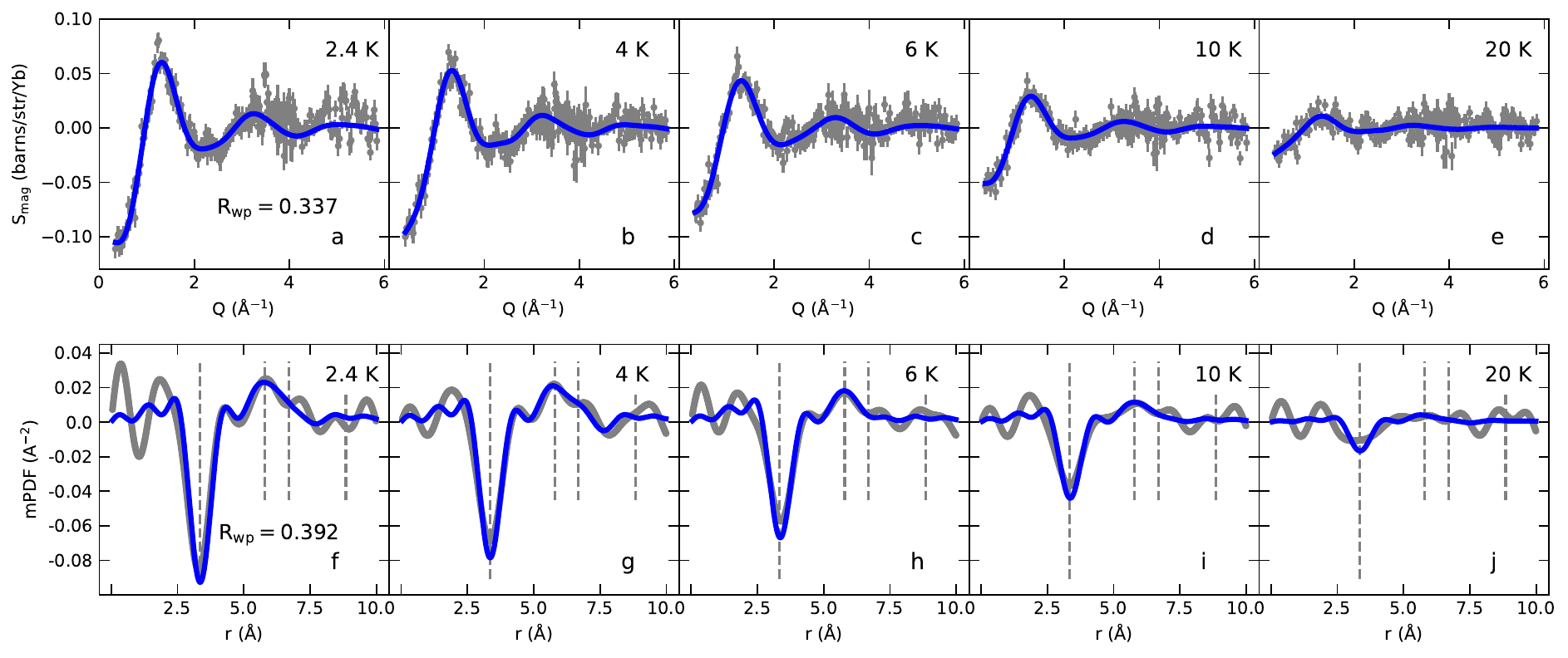}
	\caption{\label{fig:big1} (a-e) Diffuse magnetic scattering in \nyo\ at temperatures between 2.4~K and 20~K. The gray symbols represent the data and the blue curves are calculated patterns from RMC refinements using a 3D Heisenberg model. (f-j) Corresponding observed (gray) and calculated (blue) mPDF patterns obtained via Fourier transformation of the data and calculations shown in (a-e). Dashed vertical lines show the Yb-Yb near-neighbor distances. Negative (positive) peaks correspond to antiferromagnetic (ferromagnetic) correlations.}
\end{figure*}

The total scattering structure function $S(Q)$ is shown for \nyo\ at 40~K and 2.4~K in Fig.~\ref{fig:diff-struc}. Small but systematic differences exist between the data sets due to the development of magnetic correlations as the temperature is lowered. The magnetic scattering signal, isolated by subtraction of the 40-K data set from all lower temperatures (see Methods section), is shown for 2.4~K as the gray curve in Fig.~\ref{fig:diff-struc}. Zoomed-in views for 2.4~K and all other temperatures are given in Fig.~\ref{fig:big1}(a-e).
No sharp magnetic Bragg peaks are seen; instead, we observe structured diffuse scattering, a clear indication of magnetic short-range order (SRO) in the system. We infer these short-range correlations to be antiferromagnetic given the absence of a significant increase as $Q$ goes to zero, which is typically a hallmark of ferromagnetic correlations. The diffuse scattering is most prominent at 2.4~K and gradually weakens with increasing temperature as thermal fluctuations disrupt the SRO. Nevertheless, clear features are present even at 20~K, indicating that short-range correlations survive up to at least this temperature. This is consistent with earlier experimental results using other techniques~\cite{borde;np19,ding;prb19,ranji;prb19}.

We generated the corresponding real-space mPDF patterns from the diffuse magnetic scattering data at each temperature, which we display as the gray curves in Fig.~\ref{fig:big1}(f-j). The vertical dashed lines mark the Yb-Yb distances based on the known crystal structure of \nyo. The most prominent feature in the mPDF is the large negative peak centered on the nearest neighbor (NN) Yb-Yb distance, confirming the antiferromagnetic nature of the short-range correlations inferred from the diffuse scattering data in reciprocal space. The oscillations seen at distances shorter than the NN distance are nonphysical features resulting from noise in the diffuse scattering signal. Focusing on the mPDF pattern at 2.4~K in Fig.~\ref{fig:big1}(f), we note a positive peak centered on the next nearest neighbor (NNN) distance, revealing net ferromagnetic correlations between the second nearest neighbors. At further distances, the features in the mPDF are less well defined, as expected given the short-range nature of the magnetic correlations. With increasing temperature, the mPDF signal becomes weaker and restricted to shorter distances in real space. At 20~K, only a small dip at the NN distance remains, indicating that further-neighbor correlations are no longer observable at that temperature. The ease with which the nature of the local magnetic correlations and their spatial extent in real space can be determined by direct inspection of the mPDF pattern highlights the value of this approach for studying short-range magnetism.

To establish a more detailed understanding of the magnetic SRO in \nyo, we performed a series of RMC and small-box refinements to the diffuse scattering data using spin models with different lattice and spin dimensionalities. In the context of this work, ``model'' refers to a real-space configuration of spins, not to a model Hamiltonian with various types of exchange terms. The model parameters optimized in the refinements include the spin orientations and, in the case of small-box refinements, a correlation length. We began with the most general possible spin model: Heisenberg spins placed in a three-dimensional (3D) volume of the crystal structure of dimensions 20.07~\AA\ along the \textit{a} axis, 20.07~\AA\ along \textit{b}, and 16.46~\AA\ along \textit{c} (i.e. $6 \times 6 \times 1$ unit cells using the hexagonal setting). This model, which we refer to as the 3D Heisenberg model, allows spin correlations both within and between layers. For each temperature, we performed 100 refinements and averaged the calculated diffuse scattering patterns, resulting in the blue curves overlaid on the data in Fig.~\ref{fig:big1}(a-e). The corresponding calculated mPDF patterns are likewise shown in blue in Fig.~\ref{fig:big1}(f-j). The RMC-generated patterns match the experimental data well. For 2.4~K, the goodness-of-fit metric $R_{wp} = \sqrt{\sum_i\left[\left(y_i^{obs} - y_i^{calc}\right)/\sigma_i\right]^2 / \sum_i\left[y_i^{obs}/\sigma_i\right]^2}$ is 0.337 and 0.392 for the scattering and mPDF data, respectively. The full data range shown in the figures was used for calculating $R_{wp}$.

\begin{figure}
	\includegraphics[width=80mm]{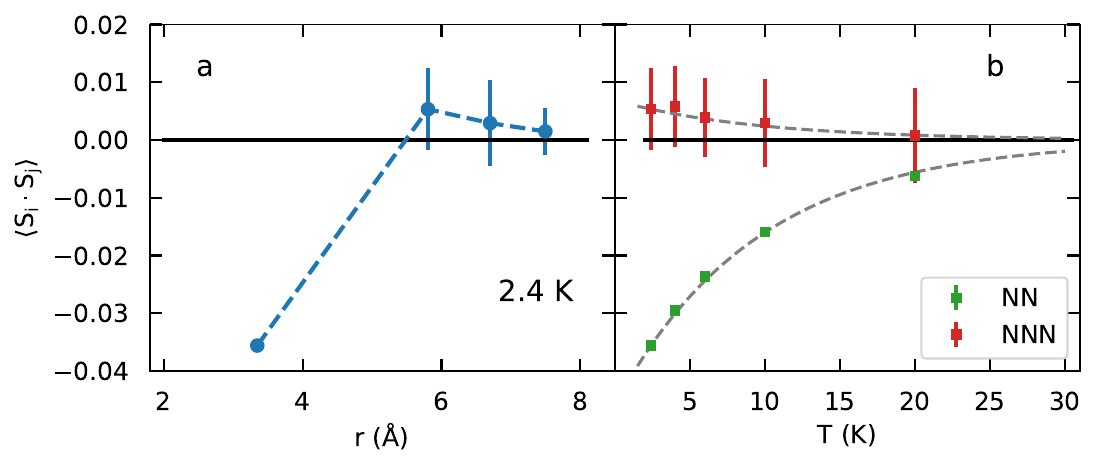}
	\caption{\label{fig:scf} (a) Spin correlation function up to the fourth nearest neighbor extracted from RMC refinements to the data collected at 2.4~K. Error bars represent one standard deviation based on the distribution of spin correlations determined from the RMC refinements. (b) The RMC-calculated NN and NNN spin correlation function in \nyo\ as a function of temperature. They gray dashed curves are guides to the eye.}
\end{figure}

From the 100 spin configurations produced by the RMC algorithm at each temperature, we computed the average spin correlation function $\langle \mathbf{S}_i \cdot \mathbf{S}_j\rangle$ up to fourth nearest neighbors. Including 100 configurations ensured that the variation in the calculated spin correlations for each coordination shell was dominated by the uncertainty in the data rather than statistical noise in the RMC output. The correlation function is displayed for 2.4~K in Fig.~\ref{fig:scf}(a).
As expected based on the mPDF data, the strongest correlation is the negative (i.e. antiferromagnetic) NN correlation, with the NNN correlation being positive (i.e. ferromagnetic) and much weaker. We note that the standard deviation of the NNN correlation, represented by the error bars in Fig.~\ref{fig:scf}, just overlaps with zero; based on the clear positive feature at the NNN distance in the mPDF data, however, we judge this positive correlation to be meaningful. The further neighbor correlations are likewise slightly positive but with larger uncertainties that overlap more significantly with zero, making it difficult to conclude with confidence anything more than the third and fourth neighbor correlations, if nonzero, are quite small. The temperature dependence of the NN and NNN correlations are shown in Fig.~\ref{fig:scf}(b), demonstrating the expected decrease as the temperature rises. Extrapolating the trend to higher temperature, we estimate the NN correlations to be negligible slightly above 30~K. Extrapolating in the opposite direction, the magnitude of the correlations can be expected to grow steeply as the temperature is lowered further.

To test the importance of interlayer magnetic correlations for describing the data, we performed RMC refinements using Heisenberg spins as before, but with the model restricted to a single 2D slice of the crystal. This eliminates all interplanar magnetic correlations from the model. Averaging together 100 such refinements leads to a fit that is practically indistinguishable from the 3D Heisenberg model; the value of $R_{wp}$ for the diffraction data at 2.4~K is 0.338, compared to 0.337 for the 3D model. We can therefore state with confidence that interlayer correlations are not necessary to describe the data fully.

We can place further constraints on the magnetic SRO through small-box modeling or RMC refinements using spin models with fewer degrees of freedom. With that in mind, we performed RMC refinements with distinct spin models based on XY spins constrained to lie in the \textit{ab} plane and Ising spins pointing along \textit{c}, along with small-box modeling of 120$^{\circ}$ SRO and stripe SRO. In all cases, we used a two-dimensional (2D) slice of the crystal. Finally, we also attempted to model the data using a single pair of antiferromagnetically aligned spins.

\begin{figure*}
	\includegraphics[width=160mm]{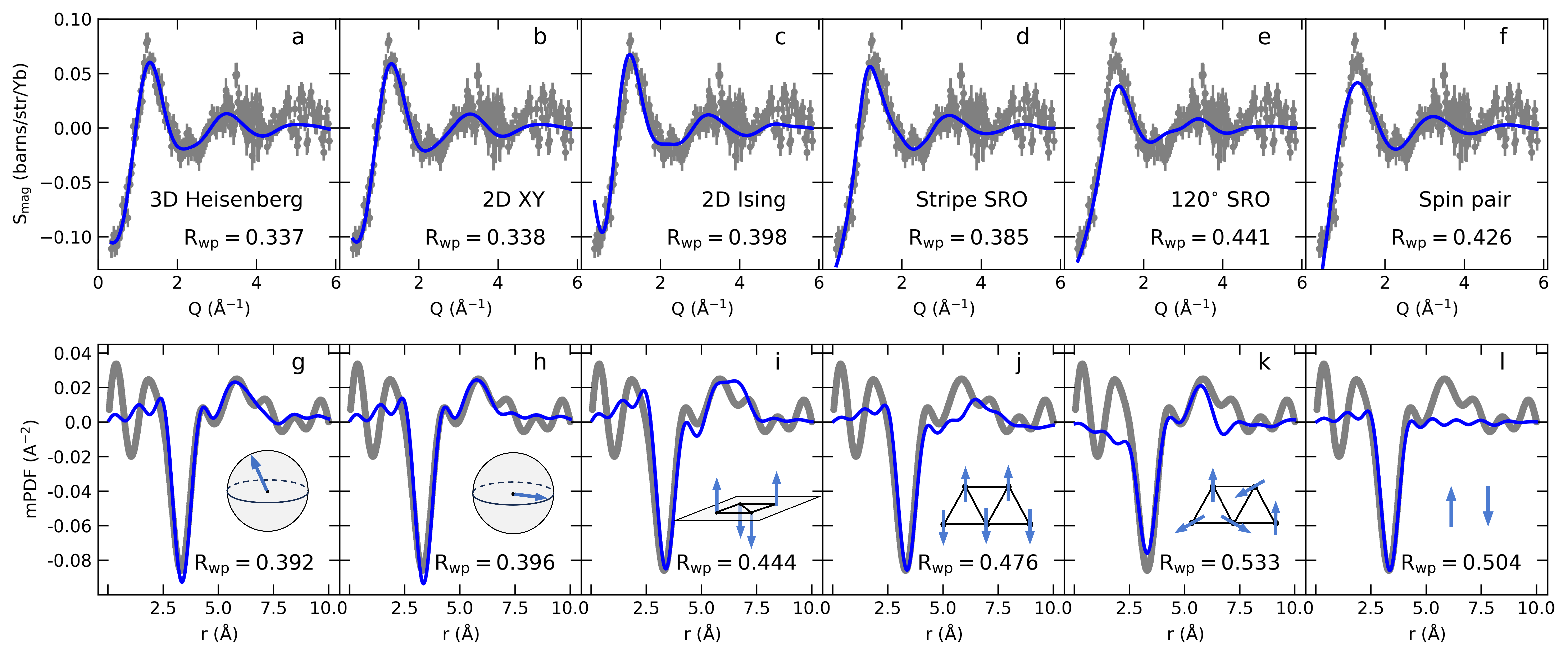}
	\caption{\label{fig:big2} (a-f) Measured (gray symbols) and calculated (blue curves) diffuse magnetic scattering in \nyo\ at 2.4~K. Each panel corresponds to a different model labeled in the figure and described in the main text. The goodness-of-fit metric $R_{wp}$ is also shown. (g-l) Equivalent results in real space. In some cases, the shortcomings of the model are highlighted more clearly in real space than reciprocal space. Insets: Schematic representations of the spin models used for the fits.}
\end{figure*}
The fitting results for each model at 2.4~K are shown in both reciprocal space and real space Fig.~\ref{fig:big2}, along with their respective $R_{wp}$ values. We see that the fit quality of the 2D XY model is virtually identical to that of the 3D and 2D Heisenberg models, indicating that this model, which contains far fewer degrees of freedom than the Heisenberg models, is equally valid for describing the data. In contrast, the other spin models all produce worse fits to the data, as indicated by the larger $R_{wp}$ values and confirmed by visual inspection. Notably, the \onetwenty\ and stripe orders, which feature in the theoretically calculated phase diagram for triangular lattice antiferromagnets~\cite{zhu;prl18,zhang;prb22}, are not consistent with the data for \nyo. Likewise, Ising spins are unable to provide as accurate a description of the data as XY or Heisenberg spins. Interestingly, the model consisting of a single pair of spins manages to capture the major features of the diffuse scattering fairly well despite its simplicity [see Fig.~\ref{fig:big2}(e)], but ultimately misses the subtler features arising from longer-range correlations.

Of the various spin models tested, the 3D Heisenberg, 2D Heisenberg, and 2D XY models provide the best fits to the data at 2.4~K. These models should all be considered equally compatible with the data, since no meaningful difference in fit quality exists among them. Considering that the 2D XY model contains significantly fewer degrees of freedom for RMC optimization than the 3D and 2D Heisenberg models, we suggest that this may be the most appropriate framework in which to consider the ground-state magnetic properties of \nyo. A scenario of short-range correlated XY spins is consistent with previous inelastic neutron scattering work~\cite{borde;np19}. That inter-layer correlations are not needed to describe the data is a positive result for the prospect of \nyo\ possessing a QSL ground state, since inter-layer interactions are typically considered detrimental to QSL stabilization in triangular lattices.

Finally, we emphasize that it can be advantageous to inspect the fits in both reciprocal and real space, since deficiencies of any given model may be more evident in one space than the other. For example, the seemingly reasonable fit of the simplistic spin-pair model to the diffuse scattering in Fig.~\ref{fig:big2}(e) is revealed to be much worse in real space [Fig.~\ref{fig:big2}(j)], where it (by design) completely misses the NNN correlation. In general, interpreting model misfits in real space is more direct and intuitive than doing so in reciprocal space.

\section{Conclusion}
Based on the experimentally measured diffuse magnetic scattering and mPDF patterns obtained from \nyo, we have shown that short-range antiferromagnetic correlations exist in \nyo\ at 2.4~K on a length scale of approximately 1~nm (i.e. three Yb-Yb nearest neighbor distances). These correlations become weaker and more short-ranged as the temperature increases, with antiferromagnetic correlations between nearest neighbor Yb moments persisting to approximately 30~K. The data can be equally well described using a 3D Heisenberg, 2D Heisenberg, or 2D XY model, while models built from Ising spins, stripe SRO, or \onetwenty\ SRO provide worse fits to the data. Our findings are consistent with the possibility of a QSL ground state in \nyo\ based on the fact that no long-range or incipient long-range order is observed and that interlayer correlations are not required for a full description of the data. These findings significantly increase our understanding of the low-temperature magnetism in this promising system. In addition, the simultaneous analysis of magnetic scattering data in both reciprocal and real space demonstrated here provides a valuable template for future studies of geometrically frustrated magnets and other systems with correlated paramagnetic states.

\textbf{Acknowledgements}

B.A.F. was supported by the U.S. Department of Energy, Office of Science, Basic Energy Sciences (DOE-BES) through Award No. DE-SC0021134. K.B.N. and C.Z.S. were supported by the College of Physical and Mathematical Sciences at Brigham Young University. S.D.W. and M.B. acknowledge financial support from DOE-BES Division
of Materials Sciences and Engineering under Grant No.
DE-SC0017752. We thank the Institut Laue-Langevin for use of its neutron instrumentation.

\end{document}